# Generalization of abelian gauge symmetry and the dark matter and energy problem

Nikolay P. Tretyakov[*][†] and Alexandre Ya. Terletsky[‡]

A commutative generalization of the $U(1)$ gauge symmetry group is proposed. The two-parametric family of two-connected abelian Lie groups is obtained. The necessity of existence of so-called imaginary charges and electromagnetic fields with negative energy density (dark photons) is derived. The possibilities when the overall Lagrangian represents a sum or difference of two identical Lagrangians for the visible and hidden sectors (i.e. copies of unbroken $U(1)$) are ruled out by the extended symmetry. The distinction between the two types of fields resides in the fact that for one of them current and electromagnetic kinetic terms in Lagrangians are identical in sign, whereas for another type these terms are opposite in sign. As a consequence, and in contrast to the common case, like imaginary charges attract and unlike charges repel. Some cosmological issues of the proposed hypothesis are discussed. Particles carrying imaginary charges ("allotons") are proposed as dark matter candidates. Such a matter would be imaginary charged on a large scale for the reason that dark atoms would carry non-compensated charges. Consequently, there exist (dark) electromagnetic fields with negative energy density on cosmological scales (the reviving of the idea of Faradayan cosmology). This leads to the hypothesis that the modern state of the Universe is radiation-dominated by dark photons with negative energy density that is the source of the observed late-time cosmological acceleration. This provides an explanation for the small value of the cosmological constant as a renormalized vacuum energy.

PACS numbers: 11.15.-q, 14.70.-e, 95.35.+d, 95.36.+x

## 1. INTRODUCTION

The idea of interactions (even long-range $U(1)$ interactions) in the dark sector is not new [1]. A new kind of photon, which couples to dark matter but not to ordinary matter, has been recently proposed by L. Ackerman et al. [2]. As mentioned in [3], an attractive non-gravitational force between DM concentrations is not only well-motivated theoretically, it may resolve some discomforts with conventional ΛCDM. DM could be weakly coupled to long-range forces, which might be related to dark energy. One difficulty with the latter is that such forces are typically mediated by scalar fields, and it is hard to construct natural models in which the scalar field remains massless (to provide a long-range force) while interacting with the DM at an interesting strength [2]. The authors of [2] point out that the dark photon comes from gauge symmetry, just

---

[*] Department of Applied Mathematics, State Social University of Russia, Moscow, Russia
[†] Corresponding author, e-mail: trn12@smtp.ru
[‡] Department of Experimental Physics, Peoples' Friendship University of Russia, Moscow, Russia

like the ordinary photon, and its masslessness is therefore completely natural. The proposed Lagrangians for the dark sector are of the type

$$L^{Dark} = \bar{\chi}\left(i\slashed{D} + m_\chi\right)\chi - \frac{1}{4}F_{\mu\nu}F^{\mu\nu}, \qquad (1)$$

where $D_\mu = \partial_\mu - igA_\mu$ and $F_{\mu\nu}$ is the field-strength tensor for the dark photons. In essence, proposed here and in some analogous works are several exact copies of $U(1)$. But, given the importance of symmetries in the SM, could we get an understanding as to where the additional $U(1)$ comes from? This is our main motivation of the present work.

In general, most SM extensions include hidden sectors, i.e. sectors which couple only very weakly, typically gravitationally, to the SM fields and these hidden sectors (if any) contain undoubtedly gauge groups as well. We suppose it is from a mechanism of extension of known groups. Because of this, simple copies of $U(1)$ are of course unlikely, since any true generalization is perceived to gain a more penetrating insight into the symmetry. It should be remembered that "More is different" [4]. The philosophy of symmetry implies that entities involved are at once similar and dissimilar to one another. Dissimilarity (even contrast) is of the same importance as similarity.

While on the subject of extension of electromagnetism, one should comply with some obvious rules. Firstly, such a simple thing as electromagnetism must certainly come from an abelian gauge group. Secondly, this group must be a non-trivial extension of the common $U(1)$ (not merely a direct sum $U(1) \oplus U'(1)$). Thirdly, the new sector, inhabited by particles charged under the new symmetry, should represent a symmetrical image of the known sector, with some properties complementing each other.

Two decades ago Ya.P. Terletsky advanced the hypothesis about the existence of so-called imaginary charges (IC) and electromagnetic fields with negative energy density ("minus-fields") [5]. The term "imaginary charge" is due to the formal substitution of imaginary values $\pm iq$ for real values $\pm q$ into the Coulomb law. In that case like charges would attract and unlike charges would repel. It is precisely this property that represents the basic physical distinction of IC from common charges, while their representation as imaginary quantities is nothing more than a mathematical tool.

The phenomenological deduction of equations for IC, given in [5], is just based on such a substitution of imaginary charges and fields into the Maxwell equations and the Lorentz force law. One possibility for such equations is the following:

$$
\begin{aligned}
&div\mathbf{E} = 4\pi\rho_e; &&div\,\mathbf{e} = -4\pi\rho_m; \\
&div\mathbf{B} = 0; &&div\,\mathbf{b} = 0; \\
&rot\mathbf{B} = \frac{4\pi}{c}\mathbf{j}_e + \frac{1}{c}\frac{\partial \mathbf{E}}{\partial t}; &&rot\,\mathbf{b} = -\frac{4\pi}{c}\mathbf{j}_m + \frac{1}{c}\frac{\partial \mathbf{e}}{\partial t} \\
&rot\mathbf{E} = -\frac{1}{c}\frac{\partial \mathbf{B}}{\partial t}; &&rot\,\mathbf{e} = -\frac{1}{c}\frac{\partial \mathbf{b}}{\partial t}
\end{aligned}
\qquad(2)
$$

$$
\mathbf{F}_{Lor} = \rho_e \mathbf{E} + \frac{1}{c}[\mathbf{j}_e \mathbf{B}] + \rho_m \mathbf{e} + \frac{1}{c}[\mathbf{j}_m \mathbf{b}].
$$

Here $\mathbf{E}, \mathbf{B}, \mathbf{e}, \mathbf{b}$ are common and "minus" fields respectively and $\rho_e, \mathbf{j}_e, \rho_m, \mathbf{j}_m$ represent their sources. The only peculiarity is the "wrong" sign in front of the sources $\rho_m$ and $j_m$ in the equations for "minus" fields, meanwhile the Lorentz force law is unchanged. This causes profound alterations in the physics of IC. In particular, one can see from the Coulomb law $div\,\mathbf{e} = -4\pi\rho_m$ and the electrostatic force $\mathbf{F}_{Lor} = +\rho_m \mathbf{e}$ that like charges would attract and unlike charges would repel. In a similar way equally directed currents would repel, in contrast to the common case.

A standard deduction of Poynting's theorem from (2) brings the following expressions for the energy density and the Poynting vector:

$$
w = \frac{1}{8\pi}\left(\mathbf{E}^2 + \mathbf{B}^2\right) - \frac{1}{8\pi}\left(\mathbf{e}^2 + \mathbf{b}^2\right); \quad \mathbf{S} = \frac{c}{4\pi}[\mathbf{E}\,\mathbf{B}] - \frac{c}{4\pi}[\mathbf{e}\,\mathbf{b}], \qquad(3)
$$

whence it follows that the energy density of minus-fields is negative. G.E. Marsh points out that "…the idea of negative energy states is still quite controversial and much confused in the literature. On the other hand, presently the behavior of the universe makes difficult to resist the idea that some negative energy matter (so-called dark energy) could contribute and repel the matter, creating the observed phenomena at large red shifts" [6]. At this point, there is no escape from citing A. Linde: "This removes the old prejudice that, even though the overall change of sign of the Lagrangian (i.e. both of its kinetic and potential terms) does not change the solutions of the theory, one *must* say that the energy of all particles is positive. This prejudice was so strong, that many years ago physicists preferred to quantize particles with negative energy as antiparticles with positive energy, which caused the appearance of such meaningless concepts as negative probability. We wish to emphasize that there is no problem to perform a consistent quantization of theories which describe particles with negative energy. All difficulties appear only when there exist interacting species with both signs of energy" [7].

Note that the term "imaginary charges" has been used to designate different things. Firstly there exists a well-known technique of replacing conducting surfaces by "imaginary charges" in electrostatics (the method of images). Secondly there is a notion of sources at imaginary space-time, which are called imaginary sources in short [8]. Thirdly, speculations representing gravitation as electrostatics with an imaginary charge are sometimes encountered. Our hypothesis has nothing to do with these uses of the term. This brings up the utility of a new term for particles carrying such charges. We propose the term "alloton" from the Greek $\alpha\lambda\lambda o\varsigma$ ("different, strange").

The aim of this article is to demonstrate the necessity of existence of IC and fields with negative energy density from first symmetry principles. The plan to be followed consists in finding a natural abelian extension of $U(1)$. We take as the starting point the single idea of extension to matrices with determinant equal to -1 (however, for reasons of commutativity, axes reflections would not do here).

It turns out that one way to accomplish this task is an extended understanding of the variational principle. Since equations of motion result from making the first variation of action equal to zero, the change of sign of a Lagrangian does not affect dynamical equations. It should be reminded in this connection that the variation principle would be more correctly said to be the principle of stationary (and not minimal) action and at real trajectories the action takes extreme rather than minimal values. Hence it is sufficient to require the invariance of Lagrangians up to change of sign.

The double universe model proposed by A. Linde [7] is of particular interest in this respect. This model describes two universes, X and Y, with coordinates $x_\mu$ and $y_\alpha$, respectively and with metrics $g_{\mu\nu}(x)$ and $\bar{g}_{\alpha\beta}(y)$, containing fields $\Phi(x)$ and $\Psi(y)$ with the action of the following type:

$$S = N\int d^4x \int d^4y \sqrt{g(x)}\sqrt{\bar{g}(y)} \left[\frac{M_P^2}{16\pi}R(x) + L(\Phi(x)) - \frac{M_P^2}{16\pi}R(y) - L(\Psi(y))\right]. \quad (4)$$

A novel symmetry of the action is the symmetry under the transformation mixing the fields $\Phi(x) \leftrightarrow \Psi(x), g_{\mu\nu}(x) \leftrightarrow \bar{g}_{\alpha\beta}(x)$ and under the subsequent change of the overall sign, $S \to -S$. Linde calls this the antipodal symmetry, since it relates to each other the states with positive and negative energies.

## 2. EXTENSION OF U(1)

The generalization is based on the following reasoning. The $U(1)$ group is isomorphic to the special orthogonal group $SO(2)$, which represents the group of orthogonal $2\times 2$ matrices with unit determinant:

$$t_\beta = \begin{bmatrix} \cos\alpha & \sin\alpha \\ -\sin\alpha & \cos\alpha \end{bmatrix}. \tag{5}$$

An extension to matrices with determinant equal to -1 may be performed by adding axes reflections. However, by this way one obtains the complete orthogonal group $O(2)$, which is non-commutative even in the case of two dimensions. This is inconsistent with our attempt to derive the existence of minus-fields analogically to electromagnetic ones, i.e. from a commutative group of lagrangian symmetries.

Let us note that matrices (5) are circulant ones, which is to say that they are of the form

$$\begin{bmatrix} a & b \\ pb & a \end{bmatrix}, \tag{6}$$

with $p = -1$. For any fixed value of $p$, non-degenerate matrices (6) form a commutative group. It may be symbolized as $C_p(2, R)$ or $C_p(2, C)$, i.e. $p$-circulant $2x2$ matrices with real or complex entries. In case that $p = 1$, they are designated briefly as circulant matrices.

It is possible to construct a commutative group of circulant ($p = 1$) matrices depending on one real parameter and with determinants equal to $\pm 1$, i.e. $SC_1(2, R)$. The group consists of elements of two types having the following general and infinitesimal representations:

$$t_\alpha = \begin{bmatrix} \cosh\alpha & \sinh\alpha \\ \sinh\alpha & \cosh\alpha \end{bmatrix} \approx \begin{bmatrix} 1 & 0 \\ 0 & 1 \end{bmatrix} + \alpha \begin{bmatrix} 0 & 1 \\ 1 & 0 \end{bmatrix}; \quad s_\alpha = \begin{bmatrix} \sinh\alpha & \cosh\alpha \\ \cosh\alpha & \sinh\alpha \end{bmatrix} \approx \begin{bmatrix} 0 & 1 \\ 1 & 0 \end{bmatrix} + \alpha \begin{bmatrix} 1 & 0 \\ 0 & 1 \end{bmatrix}; \tag{7}$$

$$\det t_\alpha = 1; \quad \det s_\alpha = -1.$$

As may be seen from (7), the elements $t_\alpha$ and $s_\alpha$ are different in that the unity and the generator switch places. The group multiplication table is of the form

$$t_\alpha t_\beta = t_{\alpha+\beta}; \quad t_\alpha s_\beta = s_{\alpha+\beta}; \quad s_\alpha s_\beta = t_{\alpha+\beta}. \tag{8}$$

Let us consider the group representation space as doublets of real scalar fields with the scalar product defined using an indefinite metric:

$$\phi = \begin{bmatrix} \phi_1 \\ \phi_2 \end{bmatrix}; \quad \psi = \begin{bmatrix} \psi_1 \\ \psi_2 \end{bmatrix}; \quad \phi * \psi = \phi_1\psi_1 - \phi_2\psi_2. \tag{9}$$

Then the quadratic form $\phi * \phi$ proves to be invariant under transformations $t_\alpha$ and changes sign under transformations $s_\alpha$:

$$t_\alpha \phi = \left(\hat{1} + \alpha \hat{G}\right)\phi = \begin{bmatrix} \phi_1 + \alpha \phi_2 \\ \phi_2 + \alpha \phi_1 \end{bmatrix}; \quad s_\alpha \phi = \left(\hat{G} + \alpha \hat{1}\right)\phi = \begin{bmatrix} \phi_2 + \alpha \phi_1 \\ \phi_1 + \alpha \phi_2 \end{bmatrix};$$

$$\left(t_\alpha \phi\right) * \left(t_\alpha \phi\right) = \left(\phi_1 + \alpha \phi_2\right)^2 - \left(\phi_2 + \alpha \phi_1\right)^2 = \phi_1^2 - \phi_2^2 = \phi * \phi; \quad (10)$$

$$\left(s_\alpha \phi\right) * \left(s_\alpha \phi\right) = \left(\phi_2 + \alpha \phi_1\right)^2 - \left(\phi_1 + \alpha \phi_2\right)^2 = \phi_2^2 - \phi_1^2 = -\phi * \phi; \quad \hat{G} = \begin{bmatrix} 0 & 1 \\ 1 & 0 \end{bmatrix}.$$

By analogy with the common case, let us define covariant derivatives as

$$D_\mu \phi = \partial_\mu \phi - e A_\mu \hat{G} \phi. \quad (11)$$

Then under simultaneous transformations of field variables and gradient transformations of potentials, covariant derivatives are transformed in the same way as fields:

$$t_\alpha: \quad \phi \to \phi + \alpha \hat{G} \phi; \quad A_\mu \to A_\mu + \frac{1}{e}\partial_\mu \alpha; \quad D_\mu \phi \to D_\mu \phi + \alpha \hat{G}\left(D_\mu \phi\right)$$

$$s_\alpha: \quad \phi \to \hat{G}\phi + \alpha \phi; \quad A_\mu \to A_\mu + \frac{1}{e}\partial_\mu \alpha; \quad D_\mu \phi \to \hat{G}\left(D_\mu \phi\right) + \alpha \left(D_\mu \phi\right). \quad (12)$$

Consequently, the Lagrangian

$$L = \frac{1}{2} D_\mu \phi * D^\mu \phi - \frac{m^2}{2} \phi * \phi \quad (13)$$

is invariant under transformations $t_\alpha$ and changes sign under transformations $s_\alpha$. Hence the invariance of dynamical equations under the extended abelian group of transformations is achieved by the invariance of lagrangians up to change of sign.

However the constructed symmetry breaks down on addition of the kinetic term describing free electromagnetic fields:

$$L = \frac{1}{2} D_\mu \phi * D^\mu \phi - \frac{m^2}{2} \phi * \phi - \frac{1}{4} F_{\mu\nu} F^{\mu\nu}; \quad F_{\mu\nu} = \partial_\mu A_\nu - \partial_\nu A_\mu. \quad (14)$$

The tensor $F_{\mu\nu}$ does not change under transformations (12), so under $s_\alpha$ the first two terms of the Lagrangian (14) reverse sign, whereas the last term does not change. The overall Lagrangian (14) turns out to be non-invariant, even up to change of sign!

It is not possible to restore the symmetry without incorporation of new entities. Another field $\psi$ must exist in addition to the usual field $\phi$. The field $\psi$ interacts with its own gauge field $a_\mu$, but with the same values of the constants $e$ and $m$. However the kinetic terms of the fields $A_\mu$ and $a_\mu$ are opposite in sign:

$$L = \frac{1}{2}\left(\partial_\mu\phi - eA_\mu \hat{G}\phi\right) * \left(\partial^\mu\phi - eA^\mu \hat{G}\phi\right) - \frac{m^2}{2}\phi * \phi +$$
$$\frac{1}{2}\left(\partial_\mu\psi - ea_\mu \hat{G}\psi\right) * \left(\partial^\mu\psi - ea^\mu \hat{G}\psi\right) - \frac{m^2}{2}\psi * \psi - \qquad (15)$$
$$-\frac{1}{4}F_{\mu\nu}F^{\mu\nu} + \frac{1}{4}f_{\mu\nu}f^{\mu\nu}; \qquad f_{\mu\nu} = \partial_\mu a_\nu - \partial_\nu a_\mu.$$

In order to bring about invariance of the Lagrangian (15) up to change of sign, transformations $s_\alpha$ must be accompanied by mixing of the fields $\phi$ and $\psi$. Let us denote such transformations by capital letters:

$$\hat{T}_\alpha: \quad \phi \to \phi + \alpha\, \hat{G}\,\phi; \quad A_\mu \to A_\mu + \frac{1}{e}\partial_\mu\alpha; \quad \psi \to \psi + \alpha\, \hat{G}\,\psi; \quad a_\mu \to a_\mu + \frac{1}{e}\partial_\mu\alpha;$$
$$\hat{S}_\alpha: \quad \phi \to \hat{G}\psi + \alpha\,\psi; \quad A_\mu \to a_\mu + \frac{1}{e}\partial_\mu\alpha; \quad \psi \to \hat{G}\phi + \alpha\,\phi; \quad a_\mu \to A_\mu + \frac{1}{e}\partial_\mu\alpha. \qquad (16)$$

The representation of (16) in terms of matrices is of the form:

$$\hat{T}_\alpha : \begin{bmatrix}\phi\\ \psi\end{bmatrix} \to \left(\begin{bmatrix}\hat{1} & 0\\ 0 & \hat{1}\end{bmatrix} + \alpha \begin{bmatrix}\hat{G} & 0\\ 0 & \hat{G}\end{bmatrix}\right)\begin{bmatrix}\phi\\ \psi\end{bmatrix};$$
$$\begin{bmatrix}A_\mu\\ a_\mu\end{bmatrix} \to \begin{bmatrix}1 & 0\\ 0 & 1\end{bmatrix}\begin{bmatrix}A_\mu\\ a_\mu\end{bmatrix} + \frac{1}{e}(\partial_\mu\alpha)\begin{bmatrix}1\\ 1\end{bmatrix};$$
$$\hat{S}_\alpha : \begin{bmatrix}\phi\\ \psi\end{bmatrix} \to \left(\begin{bmatrix}0 & \hat{G}\\ \hat{G} & 0\end{bmatrix} + \alpha \begin{bmatrix}0 & \hat{1}\\ \hat{1} & 0\end{bmatrix}\right)\begin{bmatrix}\phi\\ \psi\end{bmatrix}; \qquad (17)$$
$$\begin{bmatrix}A_\mu\\ a_\mu\end{bmatrix} \to \begin{bmatrix}0 & 1\\ 1 & 0\end{bmatrix}\begin{bmatrix}A_\mu\\ a_\mu\end{bmatrix} + \frac{1}{e}(\partial_\mu\alpha)\begin{bmatrix}1\\ 1\end{bmatrix}.$$

It is seen that the group multiplication table is the same as (8):

$$T_\alpha T_\beta = T_{\alpha+\beta} \quad T_\alpha S_\beta = S_{\alpha+\beta} \quad S_\alpha S_\beta = T_{\alpha+\beta}, \qquad (18)$$

that is to say, we are dealing with a representation of the same group (four-dimensional at this time).

Each term of the Lagrangian (15) is invariant under transformations $T_\alpha$. As for transformations $S_\alpha$, the kinetic terms with covariant derivatives and the mass terms change sign and convert to one another, the kinetic terms of the gauge fields convert to one another without change of sign: $F_{\mu\nu} \leftrightarrow f_{\mu\nu}$, but their difference changes sign. By this means, the overall sign of the Lagrangian (15) changes.

## 3. DISCUSSION OF ALTERNATIVES

There is another possibility. The following Lagrangian may be written instead of (15):

$$L = \frac{1}{2}\left(\partial_\mu \phi - eA_\mu \hat{G}\phi\right)*\left(\partial^\mu \phi - eA^\mu \hat{G}\phi\right) - \frac{m^2}{2}\phi*\phi -$$
$$-\left[\frac{1}{2}\left(\partial_\mu \psi - ea_\mu \hat{G}\psi\right)*\left(\partial^\mu \psi - ea^\mu \hat{G}\psi\right) - \frac{m^2}{2}\psi*\psi\right] - \frac{1}{4}F_{\mu\nu}F^{\mu\nu} - \frac{1}{4}f_{\mu\nu}f^{\mu\nu}. \quad (15')$$

Whereas the Lagrangian (15) may be schematically presented as $L_1 = L_{mat} - F^2 + L'_{mat} + f^2$, the expression (15') is of the form $L_2 = L_{mat} - F^2 - L'_{mat} - f^2$. The Lagrangian (15') is completely invariant under transformations $T_\alpha$ and $S_\alpha$, without change of sign. In this schematic notation, the Lagrangian in the model of A. Linde (4) may be written as a difference of two identical Lagrangians: $L_3 = L_{mat} - F^2 - L'_{mat} + f^2$, and Lagrangians in models with hidden sectors representing copies of $U(1)$ (for instance, L. Ackerman et al. [2]) as a sum of two identical Lagrangians: $L_4 = L_{mat} - F^2 + L'_{mat} - f^2$.

As mentioned above, in order to bring about invariance of the Lagrangian (15) (i.e. $L_1$ in the schematic notation) up to change of sign, transformations $S_\alpha$ in Eq. (16) are accompanied by mixing of the fields $\phi$ and $\psi$. The question arises as to whether another choice might provide invariance of $L_3$ and/or $L_4$? Let us denote possible variants as A, B, C, and D:

A. $T_\alpha : (\Phi, \Psi) \to (\Phi, \Psi); F \to F; f \to f; \quad S_\alpha : (\Phi, \Psi) \to (\Psi, \Phi); F \to f; f \to F;$

B. $T_\alpha : (\Phi, \Psi) \to (\Psi, \Phi); F \to f; f \to F; \quad S_\alpha : (\Phi, \Psi) \to (\Phi, \Psi); F \to F; f \to f;$

C. $T_\alpha : (\Phi, \Psi) \to (\Psi, \Phi); F \to f; f \to F; \quad S_\alpha : (\Phi, \Psi) \to (\Psi, \Phi); F \to f; f \to F;$

D. $T_\alpha : (\Phi, \Psi) \to (\Phi, \Psi); F \to F; f \to f; \quad S_\alpha : (\Phi, \Psi) \to (\Phi, \Psi); F \to F; f \to f.$

The variant A represents (16) where common transformations $T_\alpha$ are not accompanied by mixing of fields while "imaginary" transformations $S_\alpha$ mix fields. The variant B represents the reverse case where only $T_\alpha$ are accompanied by mixing of fields. In the variant C both $T_\alpha$ and $S_\alpha$ mix fields. Finally, none of the transformations mix fields in the variant D. The point is that there are no other possibilities apart from A, B, C, and D, since any mixing of fields must be inevitably accompanied by mixing of potentials $A_\mu \leftrightarrow a_\mu$ and, consequently, $F_{\mu\nu} \leftrightarrow f_{\mu\nu}$ (otherwise covariant derivatives would be non-invariant, even up to change of sign).

Let us remind that "imaginary" transformations $S_\alpha$ reverse signs of material lagrangians $L_{mat}$, $L'_{mat}$ and both $T_\alpha$ and $S_\alpha$ keep signs of field-strength tensors $F, f$. This game of signs leads to the conclusion that the only valid possibility is the variant A applied to $L_1$ or $L_2$. There is no other way in which a Lagrangian can be invariant, even up to change of sign. By way of example, let us examine what happens to the Lagrangian $L_3$ under the transformations. In case of A, it is invariant under $T_\alpha$ and converts to $-L'_{mat} - f^2 + L_{mat} + F^2$, which represents neither $+L_3$ nor $-L_3$, under $S_\alpha$. Analogically, in cases of B and C it converts to $-L_3$ under $T_\alpha$, while $S_\alpha$ represents a problem once again. Finally, in case of D, it is invariant under $T_\alpha$ and converts to $-L_{mat} - F^2 + L'_{mat} + F^2$ under $S_\alpha$. The latter expression is neither $+L_3$ nor $-L_3$.

So, the possibilities when the overall Lagrangian represents a sum ($L_4$) or difference ($L_3$) of two identical Lagrangians for the visible and hidden sectors are ruled out by the extended symmetry. The reason is that this group imposes more rigid restrictions on the structure of Lagrangian's terms than the common $U(1)$. Indeed, now that we have a two-connected group, we need to ensure invariance (up or not to change of sign) under both types of transformations $T_\alpha$ and $S_\alpha$.

## 4. COMPLEX SCALAR FIELD REPRESENTATION

A more conventional representation is constructible. The substitution when $e \to iq; \phi_1 \to \tilde{\phi}_1; \phi_2 \to i\tilde{\phi}_2$ (imaginary charges!), transforms the Lagrangian (13) to

$$L = \left(\partial_\mu \Phi - iqA_\mu \Phi\right)\left(\partial^\mu \overline{\Phi} + iqA^\mu \overline{\Phi}\right) - m^2 \Phi \overline{\Phi}; \quad \Phi = \frac{1}{\sqrt{2}}\left(\tilde{\phi}_1 + i\tilde{\phi}_2\right); \quad \overline{\Phi} = \frac{1}{\sqrt{2}}\left(\tilde{\phi}_1 - i\tilde{\phi}_2\right), \quad (19)$$

which represents a common Lagrangian for complex scalar fields $\Phi$. Then the transformations (10) take the form (with $i\beta$ in place of $\alpha$):

$$t_\alpha: \quad \Phi \to \Phi + i\beta\Phi = e^{i\beta}\Phi; \quad \overline{\Phi} \to \overline{\Phi} - i\beta\overline{\Phi} = e^{-i\beta}\overline{\Phi}$$

$$s_\alpha: \quad \Phi \to \Phi + i\beta\Phi = e^{i\beta}\Phi; \quad \overline{\Phi} \to -\overline{\Phi} + i\beta\overline{\Phi} = -e^{-i\beta}\overline{\Phi} \quad (20)$$

and in both cases $A_\mu \to A_\mu + \frac{1}{e}\partial_\mu \alpha = A_\mu + \frac{1}{iq}\partial_\mu(i\beta) = A_\mu + \frac{1}{q}\partial_\mu \beta$. It can be shown that in this case, too, covariant derivatives

$$D_\mu \Phi = \left(\partial_\mu \Phi - iqA_\mu \Phi\right); \quad D^\mu \overline{\Phi} = \left(\partial^\mu \overline{\Phi} + iqA^\mu \overline{\Phi}\right) \tag{21}$$

are transformed according to (20), that is, in the same manner as the fields $\Phi, \overline{\Phi}$ and consequently the Lagrangian (19) is invariant under transformations $t_\alpha$ and changes sign under $s_\alpha$.

The transformations (20) in terms of real and imaginary parts of $\Phi, \overline{\Phi}$ look like (writing $\alpha$ for $\beta$):

$$t_\alpha : \begin{bmatrix} \tilde{\phi}_1 \\ \tilde{\phi}_2 \end{bmatrix} \to \begin{bmatrix} \cos\alpha & -\sin\alpha \\ \sin\alpha & \cos\alpha \end{bmatrix} \begin{bmatrix} \tilde{\phi}_1 \\ \tilde{\phi}_2 \end{bmatrix}; \quad s_\alpha : \begin{bmatrix} \tilde{\phi}_1 \\ \tilde{\phi}_2 \end{bmatrix} \to \begin{bmatrix} i\sin\alpha & i\cos\alpha \\ -i\cos\alpha & i\sin\alpha \end{bmatrix} \begin{bmatrix} \tilde{\phi}_1 \\ \tilde{\phi}_2 \end{bmatrix}. \tag{22}$$

Let us rewrite (22) in infinitesimal form separating out the imaginary unit, as is the convention in the standard form of gauge transformations:

$$\varphi_A \to \varphi'_A = \varphi_A + i\alpha_j (T_j)_{AB} \varphi_B; \quad j = 1,\cdots,K; \; A,B = 1,\cdots,R. \tag{23}$$

Here $K$ is the dimension of the gauge group (1 in this case), $R$ is the dimension of the representation (1 in this case), $(T_j)_{AB}$ are group generators. The expressions (22) assume the form:

$$t_\alpha : \begin{bmatrix} \tilde{\phi}_1 \\ \tilde{\phi}_2 \end{bmatrix} \to \left\{ \begin{bmatrix} 1 & 0 \\ 0 & 1 \end{bmatrix} + i\alpha \left( -i \begin{bmatrix} 0 & -1 \\ 1 & 0 \end{bmatrix} \right) \right\} \begin{bmatrix} \tilde{\phi}_1 \\ \tilde{\phi}_2 \end{bmatrix};$$

$$s_\alpha : \begin{bmatrix} \tilde{\phi}_1 \\ \tilde{\phi}_2 \end{bmatrix} \to \left\{ -i \begin{bmatrix} 0 & -1 \\ 1 & 0 \end{bmatrix} + i\alpha \begin{bmatrix} 1 & 0 \\ 0 & 1 \end{bmatrix} \right\} \begin{bmatrix} \tilde{\phi}_1 \\ \tilde{\phi}_2 \end{bmatrix}. \tag{24}$$

It is seen once again from Eq. (24) that the elements $t_\alpha$ and $s_\alpha$ are different in that the unity and the two-dimensional rotation generator $i\varepsilon_{AB}$ switch places. The multiplication table is of the form (8).

From the preceding, it may be seen that the analogue of the Lagrangian (15) looks like:

$$L = \left(\partial_\mu \Phi - i q A_\mu \Phi\right)\left(\partial^\mu \overline{\Phi} + i q A^\mu \overline{\Phi}\right) - m^2 \Phi \overline{\Phi} +$$
$$\left(\partial_\mu \Psi - i q a_\mu \Psi\right)\left(\partial^\mu \overline{\Psi} + i q a^\mu \overline{\Psi}\right) - m^2 \Psi \overline{\Psi} - \frac{1}{4}F_{\mu\nu}F^{\mu\nu} + \frac{1}{4}f_{\mu\nu}f^{\mu\nu}. \tag{25}$$

Eq. (25) is invariant under the transformations

$$T_\alpha : \Phi \to \Phi + i\alpha\Phi; \quad \overline{\Phi} \to \overline{\Phi} - i\alpha\overline{\Phi}; \quad \Psi \to \Psi + i\alpha\Psi; \quad \overline{\Psi} \to \overline{\Psi} - i\alpha\overline{\Psi} \tag{26}$$

and changes sign under the transformations

$$S_\alpha : \Phi \to \Psi + i\alpha\Psi; \quad \overline{\Phi} \to -\overline{\Psi} + i\alpha\overline{\Psi};$$
$$\Psi \to \Phi + i\alpha\Phi; \quad \overline{\Psi} \to -\overline{\Phi} + i\alpha\overline{\Phi}; \qquad (27)$$

whereas the potentials are transformed exactly as in (17), with substitution $e \to q$. The transformations (26), (27) may be rewritten in a matrix form analogous to (17).

The Lagrangian (25) may be represented in a standard form separating out currents:

$$L = \partial_\mu \Phi \, \partial^\mu \overline{\Phi} - m^2 \Phi \overline{\Phi} - A_\mu J^\mu_{(\Phi)} + q^2 A_\mu A^\mu \Phi \overline{\Phi} +$$
$$+ \partial_\mu \Psi \, \partial^\mu \overline{\Psi} - m^2 \Psi \overline{\Psi} - a_\mu J^\mu_{(\Psi)} + q^2 a_\mu a^\mu \Psi \overline{\Psi} - \frac{1}{4} F_{\mu\nu} F^{\mu\nu} + \frac{1}{4} f_{\mu\nu} f^{\mu\nu}; \qquad (28)$$
$$J^\mu_{(\Phi)} = iq\left(\Phi \partial_\mu \overline{\Phi} - \overline{\Phi} \partial_\mu \Phi\right); \quad J^\mu_{(\Psi)} = iq\left(\Psi \partial_\mu \overline{\Psi} - \overline{\Psi} \partial_\mu \Psi\right).$$

Hence, the difference between the two types of fields resides in the fact that for one of them current and electromagnetic kinetic terms are identical in sign, whereas for another type these terms are opposite in sign. In the former case, varying with respect to field variables and currents, one obtains the conventional Maxwell equations and the Lorentz force law. Otherwise, the equations (2) with imaginary charges and negative energy density result. Let us notice that in case of the Lagrangian (15') the same equations are derivable, since there is no interaction between the two types of currents and the variation over them is performed independently. The physics does not depend on the overall sign of the Lagrangian (or part of it subject to variation) but on the relation between the signs of its terms.

For definiteness, we took $\Phi$ to be a scalar, though everything may be rewritten in terms of fermions as well.

## 5. DIFFERENT COUPLING CONSTANTS

Our model is parameterized by two numbers: m, the mass and q, the elementary charge (or the hidden fine-structure constant). In the foregoing consideration they are the duplicates of the couplings and mass scales of the visible sector. Of interest is the possibility of existence of two different sets of constants. As to masses, it is not evident how to perform an extension of this kind. Let us modify the Lagrangian (25) in the following way:

$$L = \left(\partial_\mu \Phi - i e A_\mu \Phi\right)\left(\partial^\mu \overline{\Phi} + i e A^\mu \overline{\Phi}\right) - m^2 \Phi \overline{\Phi} +$$
$$\left(\partial_\mu \Psi - i q a_\mu \Psi\right)\left(\partial^\mu \overline{\Psi} + i q a^\mu \overline{\Psi}\right) - m^2 \Psi \overline{\Psi} - \frac{1}{4} F_{\mu\nu} F^{\mu\nu} + \frac{q^2}{e^2} \frac{1}{4} f_{\mu\nu} f^{\mu\nu}, \qquad (25')$$

where $e$ and $q$ are different coupling constants. The transformations for fields do not differ from (26), (27), whereas the gauge transformations of potentials look like:

$$\hat{T}_\alpha: \quad A_\mu \to A_\mu + \frac{1}{e}\partial_\mu \alpha; \quad a_\mu \to a_\mu + \frac{1}{q}\partial_\mu \alpha; \quad F_{\mu\nu} \to F_{\mu\nu}; \quad f_{\mu\nu} \to f_{\mu\nu}; \quad (29)$$

$$\hat{S}_\alpha: \quad A_\mu \to \frac{q}{e}a_\mu + \frac{1}{e}\partial_\mu \alpha; \quad a_\mu \to \frac{e}{q}A_\mu + \frac{1}{q}\partial_\mu \alpha; \quad F_{\mu\nu} \to \frac{q}{e}f_{\mu\nu}; \quad f_{\mu\nu} \to \frac{e}{q}F_{\mu\nu}. \quad (30)$$

It is easily seen that the multiplication table (18) holds for (29), (30) and the covariant derivatives

$$D_\mu^{(e,A)}\Phi = \left(\partial_\mu \Phi - ieA_\mu \Phi\right), \; D_\mu^{(e,A)}\overline{\Phi} = \left(\partial_\mu \overline{\Phi} + ieA_\mu \overline{\Phi}\right),$$
$$D_\mu^{(q,a)}\Psi = \left(\partial_\mu \Psi - iqa_\mu \Psi\right), \; D_\mu^{(q,a)}\overline{\Psi} = \left(\partial_\mu \overline{\Psi} + iqa_\mu \overline{\Psi}\right) \quad (21')$$

are transformed under (26) - (27) and (29) – (30) as follows:

$$T_\alpha: D_\mu^{(e,A)}\Phi \to D_\mu^{(e,A)}\Phi + i\alpha D_\mu^{(e,A)}\Phi; \quad D_\mu^{(e,A)}\overline{\Phi} \to D_\mu^{(e,A)}\overline{\Phi} - i\alpha D_\mu^{(e,A)}\overline{\Phi};$$
$$D_\mu^{(q,a)}\Psi \to D_\mu^{(q,a)}\Psi + i\alpha D_\mu^{(q,a)}\Psi; \quad D_\mu^{(q,a)}\overline{\Psi} \to D_\mu^{(q,a)}\overline{\Psi} - i\alpha D_\mu^{(q,a)}\overline{\Psi}; \quad (31)$$

$$S_\alpha: D_\mu^{(e,A)}\Phi \to D_\mu^{(q,a)}\Psi + i\alpha D_\mu^{(q,a)}\Psi; \quad D_\mu^{(e,A)}\overline{\Phi} \to -D_\mu^{(q,a)}\overline{\Psi} + i\alpha D_\mu^{(q,a)}\overline{\Psi};$$
$$D_\mu^{(q,a)}\Psi \to D_\mu^{(e,A)}\Phi + i\alpha D_\mu^{(e,A)}\Phi; \quad D_\mu^{(q,a)}\overline{\Psi} \to -D_\mu^{(e,A)}\overline{\Phi} + i\alpha D_\mu^{(e,A)}\overline{\Phi}. \quad (32)$$

So, covariant derivatives are transformed in the same way as fields $\Phi, \overline{\Phi}, \Psi, \overline{\Psi}$ and thus the Lagrangian (25') is invariant under transformations $T_\alpha$ and changes sign under $S_\alpha$. An analogue of the completely invariant Lagrangian (15') with different couplings may be constructed as well:

$$L = \left(\partial_\mu \Phi - i\,e\,A_\mu\,\Phi\right)\left(\partial^\mu \overline{\Phi} + i\,e\,A^\mu\,\overline{\Phi}\right) - m^2 \Phi\,\overline{\Phi} -$$
$$- \left[\left(\partial_\mu \Psi - i\,q\,a_\mu\,\Psi\right)\left(\partial^\mu \overline{\Psi} + i\,q\,a^\mu\,\overline{\Psi}\right) - m^2 \Psi\,\overline{\Psi}\right] - \frac{1}{4}F_{\mu\nu}F^{\mu\nu} - \frac{q^2}{e^2}\frac{1}{4}f_{\mu\nu}f^{\mu\nu}. \quad (25'')$$

The Lagrangian (25'') is invariant (without change of sign) under both $T_\alpha$ and $S_\alpha$.

## 6. POSSIBILITY OF CONNECTORS

Let us discuss the possibility of mixing terms in Lagrangians describing interactions between common and imaginary charges, i.e. a connector sector linking hidden and visible sectors. It is very tempting to have such a connector, since it gives rise to new phenomenology and implications for dark matter detection possibilities and technological applications (perhaps including perpetuum mobile of the first kind!). However it is a rather delicate question, since such an interaction may lead to a breakdown of the vacuum due to negative energy fields. This is why in most models the two copies of the standard model matter fields, corresponding to positive and negative energy, interact only weakly through gravity, i.e. any connector sector is missing (or is multiplied by extremely small constants).

As mentioned above, the extended gauge group imposes more rigid restrictions on the structure of Lagrangian's terms than the common $U(1)$. Thus a most usual mixing term $F_{\mu\nu}f^{\mu\nu}$ is invariant under the transformations (29) – (30) and so it may not be incorporated into the Lagrangian (25'), as the latter must change sign under (30). Interestingly, this term may be included in the Lagrangian (25'') which must be invariant under all transformations. Most likely it is impossible to set up a quadratic term mixing $F_{\mu\nu}$ and $f^{\mu\nu}$ that would be invariant under (29) and would change sign under (30). As an example of terms of degree three and four let us cite the expressions

$$F_{\mu\nu}F^{\nu}{}_{\alpha}f^{\mu\alpha} - f_{\mu\nu}f^{\nu}{}_{\alpha}F^{\mu\alpha}; \qquad \varepsilon_{\mu\nu\alpha\beta}\varepsilon_{\lambda\delta\psi\chi}\left(F^{\mu\lambda}F^{\nu\delta}f^{\alpha\beta}f^{\psi\chi} - f^{\mu\lambda}f^{\nu\delta}F^{\alpha\beta}F^{\psi\chi}\right). \qquad (33)$$

It can also be seen that the mixing kinetic and Yukawa-like terms are invariant under $T_\alpha$ and change sign under $S_\alpha$:

$$L_{\text{int}} = \gamma\left(D_\mu^{(e,A)}\Phi\, D^{(q,a)\mu}\overline{\Psi} + D_\mu^{(e,A)}\overline{\Phi}\, D^{(q,a)\mu}\Psi\right) + \delta\left(\Phi\overline{\Psi} + \overline{\Phi}\Psi\right). \qquad (34)$$

Note that the change of sign occurs in a non-trivial way, since mutually conjugated terms in (34) switch places under transformations $S_\alpha$.

## 7. THE U(1) EXTENSION: GENERAL CONSIDERATION

The above-described extension of the $U(1)$ group is limited to the two possible values of the parameter in circulant matrices (6): $p = \pm 1$. Besides, the presentation is mostly restricted to the infinitesimal case. In this section let us develop a more systematic approach. We start with the most general form of $2 \times 2$ commutative matrices:

$$t_{ab} = \begin{bmatrix} a + rb & b \\ pb & a - rb \end{bmatrix}. \qquad (35)$$

For any fixed values of $p$ and $r$, non-degenerate matrices (35) form a commutative group. We would like to construct an one-parametric group, so $a$ and $b$ are assumed to be functions of a single parameter $\alpha(x^\mu)$ depending on space-time coordinates: $a = a(\alpha); b = b(\alpha)$. Hence, $t_{ab}$ will be denoted as $t_\alpha$. The condition $\det(t_\alpha) = 1$ imposes the relation between $a$ and $b$:

$$a^2 - (p+r^2)b^2 = 1. \tag{36}$$

In what follows, we assume $p + r^2 \neq 0$. Let us consider a quadratic form in the linear space of doublets of real scalar fields:

$$\phi * \phi = \phi_1^2 + c_{22} \phi_2^2 + 2c_{12} \phi_1 \phi_2. \tag{37}$$

It follows from the requirement of invariance of (37) under transformations $t_\alpha$: $(t_\alpha \phi) * (t_\alpha \phi) = \phi * \phi$, that $c_{12} = -r/p$; $c_{22} = -1/p$ and the scalar product takes the form

$$\phi * \phi = \phi_1^2 - \frac{1}{p}\phi_2^2 - \frac{2r}{p}\phi_1 \phi_2. \tag{38}$$

In order to construct matrices with determinant -1, let us arrange (35) in the form of a sum

$$t_\alpha = a \begin{bmatrix} 1 & 0 \\ 0 & 1 \end{bmatrix} + b \begin{bmatrix} r & 1 \\ p & -r \end{bmatrix}, \tag{39}$$

where the first matrix (unity) represents $t_\alpha$ with $a = 1, b = 0$, and the second matrix

$$\hat{g} = \begin{bmatrix} r & 1 \\ p & -r \end{bmatrix} \tag{40}$$

represents $t_\alpha$ with $a = 0, b = 1$. The matrix $\hat{g}$ can be considered as a "generator", even though the relation (39) is not infinitesimal. We will seek for matrices with determinant -1 in the form $s_\alpha = k \hat{g} t_b$, where $k$ is a constant. Requiring $\det(s_\alpha) = -1$, one obtains

$$s_\alpha = \sqrt{p+r^2} \begin{bmatrix} b + r\dfrac{a}{p+r^2} & \dfrac{a}{p+r^2} \\ p\dfrac{a}{p+r^2} & b - r\dfrac{a}{p+r^2} \end{bmatrix}. \tag{41}$$

It can be easily verified (taking into account (36)) that the quadratic form (38) changes sign under transformations (41): $(s_\alpha \phi) * (s_\alpha \phi) = -\phi * \phi$.

The next step consists in introducing covariant derivatives

$$D_\mu \phi = \partial_\mu \phi - eA_\mu \hat{M} \phi = \begin{bmatrix} \partial_\mu \phi_1 - eA_\mu(x\phi_1 + y\phi_2) \\ \partial_\mu \phi_2 - eA_\mu(z\phi_1 + v\phi_2) \end{bmatrix}; \qquad \hat{M} = \begin{bmatrix} x & y \\ z & v \end{bmatrix}, \tag{42}$$

where we introduce a completely arbitrary matrix $\hat{M}$. The fields are transformed as follows:

$$t_\alpha : \phi = \begin{bmatrix} \phi_1 \\ \phi_2 \end{bmatrix} \to \begin{bmatrix} (a+rb)\phi_1 + b\phi_2 \\ pb\phi_1 + (a-rb)\phi_2 \end{bmatrix}; \quad s_\alpha : \phi = \begin{bmatrix} \phi_1 \\ \phi_2 \end{bmatrix} \to \sqrt{p+r^2} \begin{bmatrix} \left(b + \frac{r}{p+r^2}a\right)\phi_1 + \frac{a}{p+r^2}\phi_2 \\ \frac{pa}{p+r^2}\phi_1 + \left(b - \frac{r}{p+r^2}a\right)\phi_2 \end{bmatrix}, \quad (43)$$

while the gauge field is subject to the gradient transformation: $A_\mu \to A_\mu + \frac{1}{e}\partial_\mu \alpha$. Inserting the components of fields, transformed under $t_\alpha$, from (43) to (42), one obtains the first transformed component of the covariant derivative:

$$(D_\mu \phi)'_1 = (a+rb)\partial_\mu \phi_1 + b\partial_\mu \phi_2 + \phi_1(\partial_\mu a + r\partial_\mu b) + \phi_2 \partial_\mu b - eA_\mu x(a+rb)\phi_1 - eA_\mu bx\phi_2 - \\ eA_\mu pby\phi_1 - eA_\mu y(a-rb)\phi_2 - [x(a+rb)\phi_1 + bx\phi_2 + pby\phi_1 + y(a-rb)\phi_2]\partial_\mu \alpha. \quad (44)$$

Covariant derivatives are transformed in the same way as fields, i.e. (44) must be equal to

$$(a+rb)\partial_\mu \varphi_1 - eA_\mu x(a+rb)\varphi_1 - eA_\mu y(a+rb)\varphi_2 + b\partial_\mu \varphi_2 - eA_\mu bz\varphi_1 - eA_\mu bv\varphi_2. \quad (45)$$

Equating (44) and (45), one obtains (denoting $a' = \partial a/\partial \alpha$, $b' = \partial b/\partial \alpha$):

$$z = py; \quad v = x - 2yr; \quad (46)$$

$$\begin{cases} a' = (x-ry)a + y(p+r^2)b \\ b' = ya + (x-yr)b. \end{cases} \quad (47)$$

It is notable that the consideration of the second component of the covariant derivative transformed under $t_\alpha$, as well as both components transformed under $s_\alpha$, leads to the same relations (46) - (47). In other words, (46) – (47) are quite sufficient to provide correct transformations of covariant derivatives (42).

The exact solution of the system (47) with the initial conditions $a(0) = a_0; b(0) = b_0$ is as follows

$$a(\alpha) = \frac{1}{2}\left(a_0 + b_0\sqrt{p+r^2}\right)e^{(x-yr+y\sqrt{p+r^2})\alpha} + \frac{1}{2}\left(a_0 - b_0\sqrt{p+r^2}\right)e^{(x-yr-y\sqrt{p+r^2})\alpha}$$

$$b(\alpha) = \frac{1}{2\sqrt{p+r^2}}\left(a_0 + b_0\sqrt{p+r^2}\right)e^{(x-yr+y\sqrt{p+r^2})\alpha} - \frac{1}{2\sqrt{p+r^2}}\left(a_0 - b_0\sqrt{p+r^2}\right)e^{(x-yr-y\sqrt{p+r^2})\alpha}.$$

(48)

The determinant (36) must be equal to 1:

$$a(\alpha)^2 - (p+r^2)b(\alpha)^2 = \left(a_0^2 - b_0^2(p+r^2)\right)e^{2\alpha(x-ry)} = 1, \quad (49)$$

whence it follows that $x = ry$ and $a_0^2 - b_0^2(p + r^2) = 1$. This condition will be satisfied if $a_0 + b_0\sqrt{p + r^2} = u$; $a_0 - b_0\sqrt{p + r^2} = 1/u$. However, the multiplication rules (8) imply $u = 1$. The matrix $\hat{M}$ takes the form

$$\hat{M} = y \begin{bmatrix} r & 1 \\ p & -r \end{bmatrix} = y\hat{g} \qquad (50)$$

and one may put $y = 1$ without loss of generality. Thus, finally, we obtain the expressions for $a(\alpha), b(\alpha)$ as follows:

$$a(\alpha) = \frac{1}{2}\left(e^{\sqrt{p+r^2}\alpha} + e^{-\sqrt{p+r^2}\alpha}\right);$$

$$b(\alpha) = \frac{1}{2\sqrt{p+r^2}}\left(e^{\sqrt{p+r^2}\alpha} - e^{-\sqrt{p+r^2}\alpha}\right). \qquad (51)$$

To summarize, the extended commutative gauge groups consist of matrices (35) and (41) with entries that depend on one real parameter and are given by (51). The family of these Lie groups is parameterized by two parameters: $p, r$. Compactness or non-compactness depends on the values of these constants. As seen from (36), a group will be compact if $p + r^2 < 0$ and non-compact otherwise. The groups are two-connected. The Lagrangian (13) in which the scalar product and the covariant derivatives are now given by (38) and (42) respectively, is invariant under transformations $t_\alpha$ and changes sign under transformations $s_\alpha$ (accompanied by gradient transformations of $A_\mu$).

In case of $r = 0$; $p = 1$, the expressions (51) take the form $a = \cosh(\alpha)$; $b = \sinh(\alpha)$ and we come to the case (7) with covariant derivatives given by (11). In case of $r = 0$; $p = -1$, the case (22) results.

## 8. COSMOLOGICAL ISSUES

Clearly the existence of IC and minus-fields leads to important astrophysical and cosmological conclusions and could have interesting consequences for structure on all scales. We propose particles carrying imaginary charges ("allotons") as dark matter candidates. J.L. Feng points out that "…it is still not at all difficult to invent new particles that satisfy all the constraints, and there are candidates motivated by minimality, particles motivated by possible experimental anomalies, etc" [9]. Our candidate is motivated by the logic of the gauge group's natural extension. Just as common electromagnetic fields exist in order to compensate the effect of local gauge transformations, so "minus" fields are designed to compensate transformations under the extended group. As mentioned in [9], it is desirable for hidden dark matter to have naturally the correct relic density, just as in the case of WIMPs, and one way to achieve this goal would be to duplicate the couplings and mass scales of the visible sector in the hidden sector, so that the WIMP miracle is satisfied in the hidden sector. However, the validity of the common assumption that the thermal relic density of a stable particle with mass $m$ annihilating through interactions with couplings $g$ is $\Omega \approx \langle \sigma v \rangle \approx m^2/g^4$, calls for further investigation in regard to particles carrying IC. Conceivably the essentials of cosmological scenarios should be revised in the presence of fields with negative energy density and particles radiating such fields. Of course, this is beyond the scope of the present work and we perform neither estimations of relic abundance nor any other parameters fitting. Below we restrict our consideration to some discussions of astrophysical and cosmological issues of the proposed hypothesis.

In case that dark matter is really composed of particles carrying IC ("allotons"), its distribution differs essentially from ordinary matter. We suggest that dark matter is dominantly comprised of atomic bound states. Dark matter would be imaginary charged on a large scale for the reason that imaginary protons and positrons may combine into atoms (like charges attract!), but such an atom would carry a non-compensated charge. Hence the capacity of such matter to form clusters is considerably enhanced. Oppositely charged atoms would accumulate elsewhere in the Universe (unlike charges repel!). Because of this, different galaxies or clusters may be oppositely charged. Consequently, there exist (dark) electromagnetic fields with negative energy density on cosmological scales (the reviving of the idea of Faradayan cosmology [10]).

In general, such a matter cannot exist in the condensed state in the ordinary sense. It may exist either in the collapsed state (conceivably in galactic centers and other astrophysical objects) or in the form of gas clouds and balls of plasma (stars), where mutual coulomb attraction is compensated by gas pressure. Nevertheless, one might expect the existence of some exotic condensed states and phase transitions.

The existence of fields with negative energy density on cosmological scales leads to the hypothesis that the modern state of the Universe is radiation-dominated by photons with negative energy and the dark matter is in thermal equilibrium with the dark photon sea. Consequently it is possible to explain the small value of the cosmological constant as a renormalized vacuum energy. It is generally believed that $\Omega_\Lambda \approx 0.72; \Omega_M \approx 0.27; \Omega_\Lambda + \Omega_M \approx 1$. In case that minus-photons give considerable contribution to the energy balance of the Universe, it may be written as follows:

$$\Omega_M + \Omega_\Lambda - \Omega_{rad} = 1; \quad \Omega_{rad} = \left|\frac{\rho_{rad,0}}{\rho_c}\right|, \quad (52)$$

where $\rho_{rad,0}$ is the current energy density of dark radiation, $\rho_c$ is the critical density. Assuming that both energy density and pressure of minus-photons are negative:

$$\rho_{rad,0} < 0; \quad p_{rad} = \frac{\rho_{rad,0}}{3} < 0, \quad (53)$$

the covariant conservation-of-energy equation for the radiation component in a FRW cosmology and the Friedman equation for the expansion rate $a(t)$ take the form

$$\dot{\rho}_{rad} + 3\frac{\dot{a}}{a}(\rho_{rad} + p_{rad}) = 0 \Rightarrow |\rho_{rad}| = \frac{c}{a^4}$$

$$(\dot{a})^2 = \frac{8\pi}{3} G \rho_c \left(\frac{\Omega_M a_0^3}{a} + \Omega_\Lambda a^2 - \Omega_{rad}\frac{a_0^4}{a^2}\right). \quad (54)$$

Differentiating the Friedman equation with respect to time, we obtain

$$\ddot{a} = a\frac{4\pi}{3} G \rho_c \left(2\Omega_\Lambda - \Omega_M\left(\frac{a_0}{a}\right)^3 + 2\Omega_{rad}\left(\frac{a_0}{a}\right)^4\right). \quad (55)$$

It follows from Eq. (55) that under the condition $2(\Omega_\Lambda + \Omega_{rad}) > \Omega_M$, the modern state of the Universe is characterized by accelerated expansion $\ddot{a}_0 > 0$, as in ΛCDM. The difference resides in that we have the effective cosmological constant $\Omega_\Lambda - \Omega_{rad}$ instead of $\Omega_\Lambda$, as it follows from (52). The small value of the cosmological constant as a renormalized vacuum energy $\Omega_\Lambda$ may be thereby explained and dark photons could be the source of the observed late-time cosmological acceleration, saying that dark energy, or part of it could be not cosmological constant or scalar field (quintessence), but photons with negative energy instead.

REFERENCES

1. X. Calmet and S. K. Majee, Phys. Lett. B **79**, 3 (2009).
2. L. Ackerman, M. R. Buckley, S. M. Carroll, and M. Kamionkowski, Phys. Rev. D **79**, 023519 (2009).
3. G.R. Farrar, R.A. Rosen, Phys. Rev. Lett. **98**, 171302 (2007).
4. P.W. Anderson, Science **177**, 393 (1972).

5. J.P. Terletsky, Annales de la Fondation Louis de Broglie, **15**, 1 (1990).
6. G.E. Marsh, arXiv: 0809.1877 [math-ph].
7. A. Linde, Phys. Lett. B **200**, 272 (1988).
8. K. Hashimoto and J.E. Wang, arXiv:0510217v1 [hep-th].
9. J. L. Feng, Ann. Rev. Astron. Astrophys. **48**, 495 (2010).
10. N. Kaloper, A. Padilla, JCAP **10**, 023 (2009).